\documentclass{nsr}

\usepackage{amsmath,graphicx,array}
\usepackage{dcolumn,soul}%

\usepackage{amsthm}
\usepackage[figuresright]{rotating}%
\usepackage{algorithm, algorithmicx, algpseudocode}
\usepackage{listings}%
\usepackage{hyperref}

\makeatletter
\def\uns{\ifmmode\,\else$\,$\fi}%

\makeatother
 
\jvol{XX}
\jnum{X}
\jyear{Year}
\doi{10.1093/nsr/XXXX}
\received{XX XX Year}
\revised{XX XX Year}
\accepted{XX XX Year}

\usepackage{adjustbox}
\usepackage{graphicx}%
\usepackage{multirow}%
\usepackage[title]{appendix}%
\usepackage{textcomp}%
\usepackage{manyfoot}%
\usepackage{booktabs}%
\usepackage[T1]{fontenc}
\usepackage{float}
\usepackage{graphicx}
\usepackage{pdflscape}
\usepackage{pgf,pgffor}
\usepackage{textcomp}
\usepackage{threeparttable} 
\usepackage{xcolor}
\usepackage{ulem}
\usepackage{lineno}
\usepackage{caption}
\usepackage[caption = false]{subfig}
\usepackage{academicons}
\usepackage{pifont}
\usepackage{longtable}
\usepackage{cleveref}
\definecolor{orcidlogocol}{HTML}{A6CE39}
\usepackage[symbol]{footmisc}
\renewcommand{\thefootnote}{\fnsymbol{footnote}}
\makeatletter

\newcommand{\FIGU}[1] {Fig.~\ref{#1}}
\newcommand{\TAB}[1] {Table~\ref{#1}}

\newcommand{\AFF}[1]{{\foreach\d[count=\ni]in{#1}{\ifnum\ni=1\ref{\d}\else,\ref{\d}\fi}}}

\definecolor{dkblue}{RGB}{54, 86, 169}

\newcommand{\PP}{left hand}
\newcommand{\NP}{right hand}

\begin{document}

\dhead{RESEARCH ARTICLE}

\subhead{PHYSICS}

\title{Ninety percent circular polarization 
detected in a repeating fast radio burst}

\author{J.~C.~Jiang$^1$}
\author{J.~W.~Xu$^{2,3}$}
\author{J.~R.~Niu$^1$}
\author{K.~J.~Lee$^{2,1,4,*}$}
\author{W.~W.~Zhu$^{1,*}$}
\author{B.~Zhang$^{5,6,*}$}
\author{Y.~Qu$^{5,6}$}
\author{H.~Xu$^1$}
\author{D.~J.~Zhou$^1$}
\author{S.~S.~Cao$^2$}
\author{W.~Y.~Wang$^7$}
\author{B.~J.~Wang$^1$}
\author{S.~Cao$^8$}
\author{Y.~K.~Zhang$^1$}
\author{C.~F.~Zhang$^1$}
\author{H.~Q.~Gan$^1$}
\author{J.~L.~Han$^1$}
\author{L.~F.~Hao$^8$}
\author{Y.~X.~Huang$^8$}
\author{P.~Jiang$^1$}
\author{D.~Z.~Li$^9$}
\author{H.~Li$^1$}
\author{Y.~Li$^{10}$}
\author{Z.~X.~Li$^8$}
\author{R.~Luo$^{11}$}
\author{Y.~P.~Men$^{12}$}
\author{L.~Qian$^1$}
\author{J.~H.~Sun$^1$}
\author{L.~Wang$^3$}
\author{Y.~H.~Xu$^8$}
\author{R.~X.~Xu$^{2, 14}$}
\author{Y.~P.~Yang$^{13}$}
\author{R.~Yao$^1$}
\author{Y.~L.~Yue$^1$}
\author{D.~J.~Yu$^1$}
\author{J.~P.~Yuan$^{15}$}
\author{Y.~Zhu$^1$}

\affil{$^1$National Astronomical Observatories, CAS, Beijing 100101, China}
\affil{$^2$Department of Astronomy, Peking University, Beijing 100871, China}
\affil{$^3$Kavli Institute for Astronomy and Astrophysics, Peking University, Beijing 100871, China}
\affil{$^4$Beijing Laser Acceleration Innovation Center, Beijing 101400, China}
\affil{$^5$Nevada Center for Astrophysics, University of Nevada, Las Vegas NV 89154, USA}
\affil{$^6$Department of Physics and Astronomy, University of Nevada, Las Vegas NV 89154, USA}
\affil{$^7$School of Astronomy and Space Science, University of Chinese Academy of Sciences, Beijing 100049, China}
\affil{$^8$Yunnan Observatories, CAS, Kunming 650216, China}
\affil{$^9$Department of Astrophysical Sciences, Princeton University, Princeton NJ 08544, USA}
\affil{$^{10}$Purple Mountain Observatory, CAS, Nanjing 210008, China}
\affil{$^{11}$School of Physics and Electronic Engineering, Guangzhou University, Guangzhou 510006, China}
\affil{$^{12}$Max-Planck institut f\"ur Radioastronomie, Bonn 53121, Germany}
\affil{$^{13}$South-Western Institute For Astronomy Research, Yunnan University, Kunming 650504, China}
\affil{$^{14}$State Key Laboratory of Nuclear Physics and Technology, Peking University, Beijing 100871, China}
\affil{$^{15}$Xinjiang Astronomical Observatory, CAS, Urumqi 830011, China}

\authornote{\textbf{Corresponding authors.} Email: kjlee@pku.edu.cn, zhuww@nao.cas.cn, bing.zhang@unlv.edu}

\abstract[ABSTRACT]{Fast radio bursts (FRBs) are extra-galactic sources
with unknown physical mechanisms. They emit millisecond-duration radio
pulses with isotropic equivalent energy of $10^{36}$ -- $10^{41}$ ergs. This corresponds to a brightness temperature of FRB emission typically reaching the level of $10^{36}$~K, but can be as high as above $10^{40}$~K for sub-microsecond timescale structures, suggesting the presence
of underlying coherent relativistic radiation mechanisms. Polarization carries the key information to understand the physical
origin of FRBs, with linear polarization usually tracing the geometric
configuration of magnetic fields and circular polarization probing both
intrinsic radiation mechanisms and propagation effects. Here we show that
the repeating sources FRB~20201124A emits $90.9\pm 1.1\%$ circularly
polarized radio pulses. Such a high degree of circular polarization
was unexpected in theory and unprecedented in observation in the case of FRBs, since such a high degree of circular
polarization was only common among Solar or Jovian radio activities,
attributed to the sub-relativistic electrons. We note that there is no obvious correlation between the degree of circular polarization and burst fluence. Besides the high degree of circular polarization, we also detected rapid swing and orthogonal jump in the position angle of linear polarization. The detection of
the high degree circular polarization in FRB~20201124A, together with
its linear polarization properties that show orthogonal modes, place
strong constraints on FRB physical mechanisms, calling for an interplay
between magnetospheric radiation and propagation effects in shaping the
observed FRB radiation.  } 

\maketitle

\section{INTRODUCTION} Within the framework of the
magnetar engine as suggested by the detection of the Galactic FRB
20200428D\cite{2020Natur.587...59B,2020Natur.587...54C}, two major
types of models have been proposed\cite{Zhang2020Natur}: the far-away,
gamma-ray burst (GRB)-like models that invoke synchrotron maser in
highly magnetized, relativistic shocks\cite{Metzger19,Beloborodov20}
and the closer-in, pulsar-like models that invoke radio emission from
magnetospheres\cite{Kumar17,Yang18,2022ApJ...927..105W,2022SCPMA..6589511W,2022ApJ...925...53Z,2023MNRAS.522.2448Q}.
Previous observations already provided crucial
polarization properties of FRBs, including linear polarization angle
swings\cite{2019MNRAS.486.3636P,Luo2020Natur,2022Natur.609..685X,Hilmarsson2021MNRAS,Kumar2022MNRAS,2022RAA....22l4003J,2023PhRvD.108d3009K,2024ApJ...974..296F}, and
oscillations between linear and circular
polarizations\cite{2022Natur.609..685X}.

Moderate levels of circular polarization from 5\% to 57\% had been reported for
FRBs without observed repetition, i.e. for FRB~20110523A\cite{Masui2015Natur}, 20140514A\cite{Petroff2015MNRAS}, 20160102A\cite{Caleb2018MNRAS},
20180309A\cite{Oslowski2019MNRAS}, 20180311A\cite{Oslowski2019MNRAS}, 20180714A\cite{Oslowski2019MNRAS}, 20180924B\cite{Day2020MNRAS}, 20181112A\cite{Prochaska2019Sci,Cho2020ApJ}, 20190102C\cite{Day2020MNRAS}, 20190608B\cite{Day2020MNRAS},
20190611B\cite{Day2020MNRAS}, and 20191219F\cite{Mckinven2021ApJ}. Repeating FRBs mostly show low level of circular
polarization (a few percent) and only a few of them emit significant circular
polarization for a small fraction of bursts, i.e.
FRB~20190520B\cite{2023Sci...380..599A} (42\%)
FRB~20201124A\cite{2022Natur.609..685X,Hilmarsson2021MNRAS,Kumar2022MNRAS,2022RAA....22l4003J,2023PhRvD.108d3009K}
($\sim75\%$), FRB~20121102A\cite{2022SciBu..67.2398F} (64\%), and
FRB~20220912A\cite{2023ApJ...955..142Z,2023ApJ...949L...3R,2024ApJ...974..296F}
($70\%$). Among the sources, the FRB~20201124A is an active repeating FRB
source with event rate as high as 542~hr$^{-1}$ during its active window \cite{2022Natur.609..685X,2022RAA....22l4001Z,2022RAA....22l4002Z,2022RAA....22l4003J,2022RAA....22l4004N}. Previous observations have detected 1863 bursts from the source using the
Five-hundred-meter Aperture Spherical radio Telescope (FAST) in the L-band
(centered at 1.25 GHz) during one of its active periods between March and May,
2021\cite{2022Natur.609..685X}.  It entered a new period of activity on
September 21, 2021 according to the alert from the Canadian Hydrogen Intensity
Mapping Experiment (CHIME)\cite{CHIME2021ApJS}. Following the detections, we scheduled daily FAST observations to monitor the source between
September 25 and October 2,
2021\cite{2022RAA....22l4001Z,2022RAA....22l4002Z,2022RAA....22l4003J,2022RAA....22l4004N}.
Here we report the detection of highly circularly polarized (with the highest degree of
circular polarization $90.9\pm 1.1\%$) radio pulses from FRB 20201124A with FAST. The observation serves
as a touchstone for FRB radiation models and suggests extreme conditions for the
generation and propagation of FRBs in the nearby environment.

\section{RESULTS}
\begin{figure*} \centering \includegraphics[width=\linewidth]{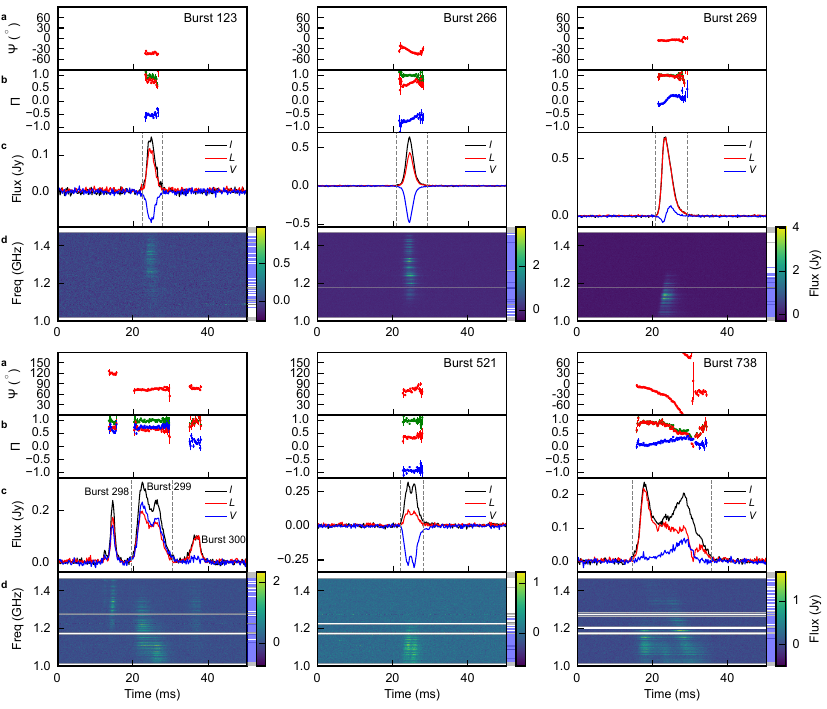}
	\caption{Polarimetric results of a selected sample of bursts with high degrees
	of circular polarization or abrupt jumps in linear polarization position
	angle. For each burst, we plot (\textbf{a}) position angle, (\textbf{b}) degree
	of linear (red), circular (blue) and total polarization (green) as a function
	of burst time, (\textbf{c}) burst profile of total intensity (black), linear
	polarization (red), and circular polarization (blue), and (\textbf{d}) dynamic
	spectra of the bursts, i.e. intensity as function of time and frequency. On the right side of each spectrum, the 
	grey shades denote the removed frequency channels affected by RFI and the 20
	MHz band edges, and the blue shades denote the frequency channels where the
	burst signal appears. The color bar manifests the total intensity in the spectrum. Here, the frequency resolution of spectra is 1.95 MHz
	per channel, and the time resolution of the light curves and spectra is 196
	$\mu$s per sample. Errorbars are for 68\% confidence level. 
    Bursts are de-dispersed with the daily average DM values published previously\cite{2022RAA....22l4001Z}, which are  $412.5\,\mathrm{pc\,cm^{-3}}$ for Bursts 123, 266 and 269, and $411.6\,\mathrm{pc\,cm^{-3}}$ for other bursts in this figure. Bursts 298, 299 and 300 are plotted in the same panel as they are very close in time.}
	\label{fig:prof_ms} \end{figure*}

In the new active window, we detected a group of highly circularly polarized 
bursts from FRB~20201124A and found the evidence of orthogonal polarization 
modes, i.e. dual linear polarization modes with orthogonal directions. In a 
total of 536 bursts with signal-to-noise ratio ($\mathrm{S/N}$) higher than 50 out of more than 800 bursts above the detection threshold $\mathrm{S/N}\ge 7$, we detected a group of 15 bursts 
with average degrees of circular polarization $\Pi_{\rm v}\equiv|V|/I>50\%$ and 
106 bursts with $\Pi_{\rm v} \ge 20\%$. A selected sample of bursts with 
significant circular polarization is shown in \FIGU{fig:prof_ms}, while other 
bursts with peak $\Pi_{\rm v}>50\%$ are shown in \FIGU{fig:highvatlas1}--\ref{fig:highvatlas4} and 
their properties are summarized in \TAB{tab:summary}.

As shown in \FIGU{fig:prof_ms}, the time-averaged $\Pi_{\rm v}$ of Bursts 266,
299 and 521 were $-71.1\%\pm 0.4\%$, $73.5\%\pm 0.7\%$ and $-90.9\%\pm 1.1\%$, respectively,
while the time-resolved $\Pi_{\rm v}$ could be even higher. The degree
of circular polarization of Burst 521 is larger than any FRB bursts
previously reported.  The two bursts have opposite handedness of circular
polarization. In Burst 299, the circular polarization is \PP, while the
\NP\ mode dominates Burst 521. The burst profile of both bursts showed
double-peak structure in total intensity $I$, linear polarization
intensity $L$, and circular polarization intensity $V$.  The linear
polarization position angle (PA) remained constant across the two peaks.

In \FIGU{fig:prof_ms}, we also show the two adjacent bursts (Bursts 298 and 300) 
of Burst 299, as they are only separated by a few tens of milliseconds from 
Burst 299. Despite the temporal proximity, these three bursts show rather 
different polarization properties. The degree of circular polarization of Burst 
298 is similar to that of Burst 299. However, the PA changed abruptly by 
approximately $60^\circ$. Bursts 299 and 300 had the same PA, while the degree 
of circular polarization dropped from $73.5\%$ to $12.2\%\pm 1.9\%$ within less than 
20 ms. In the spectral domain, the degree of linear polarization ($\Pi_{\rm 
L}\equiv L/I$) and $\Pi_{\rm v}$ remained almost constant across the observed 
bandwidth of bursts (\FIGU{fig:spectral}), i.e. there is no obvious 
polarization evolution or oscillations\cite{2022Natur.609..685X} from 1.0 GHz to 
1.4 GHz.  

Despite the similarities in Bursts 299 and 521,  the circular polarization phenomena for FRB~20201124A are of a great diversity. From the morphology perspective, not all high degree of circular polarization is associated with the double-peak pulse structure like Bursts 299 and 521. As an example, Burst 123 in \FIGU{fig:prof_ms} with $\Pi_{\rm V}=-52\%$ has a single peak pulse.
One can also note clear variations of $\Pi_{\rm v}$ across the phase in Burst 123, opposite to the case of Bursts 299 and 521. For the bursts with lower degrees of circular polarization, we have found single-peak pulses with circular polarization sign reversal (Burst 269), which is shown in \FIGU{fig:prof_ms}. We also identified abrupt $90^\circ$ jumps in PA of linear polarization. An example (Burst 738) is shown in \FIGU{fig:prof_ms}, where a $90^\circ$ jump occurs at 30 ms. One can notice that the degree of circular polarization peaked at the time of the 90$^\circ$ PA jump.

\section{DISCUSION}

The substantially high circular polarization degree observed in a few bursts provides valuable insights into the coherent radiation mechanism of FRBs. Our immediate conclusion is that the far-way GRB-like models, which invoke maser-type emission from relativistic shocks outside the magnetosphere\cite{Metzger19,Beloborodov20}, are not supported by the data. In general, circular polarization of emission can be generated either from intrinsic radiation mechanisms or through propagation effects. Both mechanisms fail to accommodate the data for GRB-like models.  
First, it has been shown \cite{2023MNRAS.522.2448Q} that the 90\% circular polarization from a bright burst cannot be generated from the intrinsic synchrotron maser (the relativistic version of cyclotron maser \cite{1985SSRv...41..215W}) model. This is because the synchrotron maser model requires ordered magnetic fields in the shock plane, and the observed radiation predominately carries high linear polarization. Circular polarization is possible with off-beam viewing geometry, but these 
high $\Pi_{\rm v}$ bursts are expected to have much lower fluxes than other bursts, inconsistent with the bright high $\Pi_{\rm v}$ bursts we detected. GRB-like models are also disfavored due to rapid swings of PA curves, which has been observed in the emission from other repeaters FRB~20180301A\cite{Luo2020Natur} and FRB~20220912A\cite{2024ApJ...974..296F}.

Our results also challenge the out-of-magnetosphere propagation models for the circular polarization. Propagation models might plausibly incorporate Faraday conversion and synchrotron or cyclotron absorption to generate a high degree of circular polarization\cite{Wang10,Lu21,2023MNRAS.522.2448Q,2022Natur.609..685X,2022NatCo..13.4382W,2022ApJ...933L...6L}. However, these models encounter difficulties in elucidating the dynamics observed in the three successive bursts within 30 milliseconds, Bursts 298, 299, and 300, as depicted in \FIGU{fig:prof_ms}. As out-of-magnetosphere propagation effects will simultaneously affect both linear and circular polarizations for all pulses, intricate conditions have to be satisfied, such that 1) the PA changed significantly while $\Pi_{\rm v}$ and $\Pi_{\rm L}$ remained nearly constant (Bursts 298 and 299), and 2) $\Pi_{\rm v}$ changed while PA being constant (Bursts 299 and 300). It is essentially impossible for an external plasma to make such abrupt changes within a timescale of 30 milliseconds. Furthermore, Bursts 521 and 123 exhibited frequency-independent degrees of circular or linear polarization, rendering the models centered solely on propagation effects increasingly difficult, because both Faraday conversion and synchrotron/cyclotron absorption processes will typically exhibit frequency dependent polarization. Furthermore, the absorption-related models require a non-negligible value of the optical depth to produce high circular polarization\cite{2021ApJ...920...46D,2023MNRAS.522.2448Q}, and thus predict a lower flux for high $\Pi_{\rm v}$ bursts. As a result, these mechanisms also suffer from the flux problem similar to the one the synchrotron maser emission model faces. 
 
The data are more consistent with pulsar-like models that invoke emission inside a magnetar magnetosphere. 
For the intrinsic radiation models, two widely discussed magnetospheric mechanisms to explain the high brightness temperatures of FRB radio emission\cite{Zhang2023RMP} involve either curvature radiation\cite{Kumar17,Yang18,2022ApJ...927..105W} or inverse Compton scattering (ICS) mechanisms\cite{Qiao98,Xu00,2022ApJ...925...53Z} of particle bunches. Regarding curvature radiation\cite{2022SCPMA..6589511W}, circular polarization can be generated, when the observer's line of sight slightly deviates from the center of radiation beam, resulting in seeing the circular motion of electrons and detecting the corresponding circular polarization.  For ICS, the scattered emission by a single particle is linearly polarized, but circular polarization can be generated as a consequence of emission from a bunch with certain geometry. In particular, 
when the line of sight mis-aligns with the direction of the ICS electron bunch, circular polarization could be detected, as different phases are superposed to the scattered radiation induced by the different parts of the bunch\cite{2023MNRAS.522.2448Q}. A high circular polarization degree can be generated slightly off-axis, even when the line of sight is still within the $1/\gamma$ relativistic beaming cone of the relativistic bunch ($\gamma$ being the Lorentz factor of the bunch). 

We noted the sign reversal of $\Pi_{\rm v}$ with an example showing in Burst 269. This phenomenon closely aligns with the predictions of both curvature radiation and ICS models, which predict the reversal of the sign of circular polarization when the line of sight sweeps across the meridian plane of the electron's trajectory, and hence, provides a strong support to the magnetospheric origin of FRB emission. In the mean time, our observations also provide constraints on these models. In general, since circular polarization is observed off-beam in both models, one should generally expect a systematically fainter population for the high $\Pi_{\rm v}$ bursts. 
We compared the fluence distributions of bursts with $\Pi_{\rm v} \ge 50\%$ and $\Pi_{\rm v} \le 50\%$ using the two-sample Kolmogorov-Smirnov test and could not find evidence that the bursts with $\Pi_{\rm v}\ge 50\%$ are systematically fainter (see Section S4 within the Supplementary Data). 
Our simulations showed that one would detect the difference at a confidence level higher than 95\%, if the energies of $\Pi_{\rm v}\ge 50\%$ bursts are reduced to 65\% of the observed values.  
Theories for FRB radiation need to address whether the models 
can reproduce the independence between the distributions of circular polarization degree and burst fluence. Another interesting constraint comes from the nearly constant degree of circular polarization as observed in Burst 521, which requires that the line of sight should maintain a nearly constant angle relative to the center of radiation beam in the entire burst. This is possible for the ICS mechanism which predicts a similar $\Pi_{\rm v}$ in a wide range of azimuthal angle with respect to a bunch\cite{2023MNRAS.522.2448Q} and may suggest a non-steady plasma flow along the magnetic field lines\cite{2022SCPMA..6589511W} or a very special geometric configuration of magnetic field, electron velocity, and line of sight for the curvature radiation model. 

Our results may be also accommodated for wave propagation within the magnetosphere. A possible explanation to the differences between Bursts 298, 299 and 300 is that the three bursts originate from different depths in a magnetosphere, and propagate different distances before escaping. Since circular polarization may be generated during wave propagation,  Burst 300 with the least amount of circular polarization might be least affected  by propagation effects among the three bursts. Another hint of magnetospheric propagation is that Bursts 738 shows a sudden 90$^\circ$ jump in the linear polarization position angle (approximately at $\sim 30$ ms in \FIGU{fig:prof_ms}). Similar features have been widely detected in pulsars, mostly in the averaged pulse profiles\cite{1984ApJS...55..247S}, but sometimes in single pulses as well\cite{Singh2024MNRAS.527.2612S}. Such orthogonal jumps can be attributed to the superposition of two linearly polarized wave modes (O-mode and X-mode) either incoherently\cite{1984ApJS...55..247S} or coherently\cite{Dyks2017MNRAS.472.4598D}. For our case, the observed PA jump occurs within a single burst, and degree of circular polarization forms a peak at the PA-jump epoch accompanied with a nearly full depolarization of linear polarization. These behaviors suggest a possible coherent superposition between the two linear polarization modes. 
It is more plausible that the two modes were excited by the same group of electrons, and the propagation effects, such as birefringence, within magnetosphere induced the 90$^\circ$ PA jump. Our results indicate that the propagation effects need also maintain a 90$^\circ$ phase difference between the two modes to ensure that linear polarization is depolarized and circular polarization dominates at the epoch of PA jump. The multi-path propagation effect\cite{Feng2022Sci...375.1266F} can cancel linear polarization and keep circular polarization components, but it is unexpected that the $\Pi_{\rm v}$ peaks at the PA jump. 

\begin{figure} \centering \includegraphics[width=\linewidth]{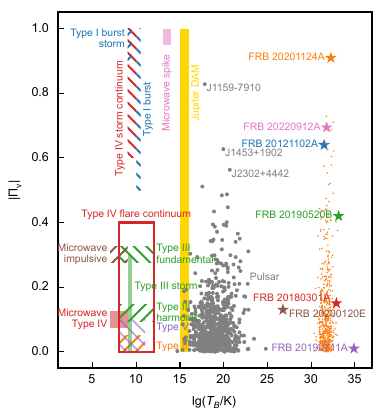}
	\caption{Astrophysical sources with significant circular polarization in the
	radio band. The x-axis is the brightness temperature. The y-axis is the degree
	of circular polarization $\Pi_{\rm v}$. For FRBs, the maximal reported
	$\Pi_{\rm v}$ is plotted with star symbols. All the bursts with $S/N\ge50$ of
	FRB~20201124A are also included (orange dots) for reader's reference. The
	details of the data sources are presented in Section~S5 of the online
	Supplementary Data.} \label{fig:docp_source} \end{figure}

FRB~20201124A belongs to a unique type of source. In \FIGU{fig:docp_source}, we summarized astrophysical radio sources known so far to exhibit significant circular polarization. One can see that FRB~20201124A stands out in the parameter space of high $\Pi_{\rm v}$ and high luminosity. The other examples with high degree of circular polarization include Type I and Type IV solar storms, as well as solar storm continua and microwave spikes, which can reach the degree of circular polarization similar to that of Burst 521. Additionally, the magnetosphere of Jupiter has been observed to generate highly circularly polarized radio emission. Unlike in the case of FRBs, these sources of high circular polarization are powered by sub-relativistic plasma. The circular polarization are mainly generated from cyclotron radiation. In contrast, the high luminosity and high degree of circular polarization in the case of FRBs demand relativistic plasma emission in a new parameter regime, which differs from all radio sources known in the past, suggesting extreme conditions for FRB generation and propagation.

In summary, the extreme and diverse polarization properties of FRB~20201124A provide constraints on the physical mechanisms generating the FRB radiation. Although a comparable degree of circular polarization has been observed in solar and Jovian radio emissions, their non-relativistic nature limits the applicability to FRB models. GRB-like models and out-of-magnetosphere propagation models are disfavored for the following reasons: 1) the models cannot account for rapid polarization changes as seen in bursts 298, 299, and 300; 2) the frequency independence of $\Pi_{\rm v}$ as observed in bursts 123 and 521 cannot be interpreted by only invoking 
Faraday conversion and synchrotron/cyclotron absorption, and 3) no strong correlation between flux and circular polarization is detected. In constrast, pulsar-like models and magnetospheric propagation models are more plausible, even though some fine-tuning of these models may be necessary. Among the known radio burst/pulse sources, the high degree of circular polarization, combined with the high brightness temperature, makes FRB radiation processes even more mysterious.

\section{METHODS}

The FAST intensively monitored the FRB~20201124A between September 25 and October 
17, 
2021\cite{2022RAA....22l4001Z,2022RAA....22l4002Z,2022RAA....22l4003J,2022RAA....22l4004N}.  
The L-band receiver covers radio frequency between 1 and 1.5~GHz, divided into 
4096 channels. The original time resolution is $49.152\,\mathrm{\mu s/sample}$.  
We used software package \texttt{TransientX}\cite{2024A&A...683A.183M} to search 
for bursts in the data, and detected more than 500 bursts with $\mathrm{S/N}>50$ 
between September 25 and 28, 2021. More details about observation and burst 
detection can be found in Section S1 within the Supplementary Data.

The signal from the noise diode was injected to the receiver as the linearly
polarized calibrator. The polarimetric calibration was performed using
\texttt{PSRCHIVE}\cite{PSRCHIVE2004PASA}. Then, the rotation measure (RM) is
measured by fitting the oscillation of Stokes $Q$ and $U$. In this article we
follow the PSR/IEEE convention on the definition of Stokes
parameters\cite{PSRCHIVE2010PASA}. Our final results shown above include both
the corrections of feed polarization response and Faraday rotation effect. The
more details about polarimetry can be found in Section S2 and S3 in the online
Supplementary Data.

\section*{DATA AVAILABILITY} Raw data are archived and available from the FAST
data center (\url{http://fast.bao.ac.cn}). Owing to the large data volume, we
encourage contacting the corresponding author and FAST data center for the data
transferring. The directly related data that support the findings of this study
can be found at the PSRPKU website (\url{https://psr.pku.edu.cn/index.php/publications/frb20201124a/}). 

CHIME/FRB Public Database can be found at \url{https://www.chime-frb.ca/repeaters/FRB20201124A}.

\section*{CODE AVAILABILITY}
The software packages to generate the results of this paper are available from \textsc{PSRCHIVE} (\url{http://psrchive.sourceforge.net}), \textsc{transientx} (\url{https://github.com/ypmen/TransientX}) and \textsc{BEAR} (\url{https://psr.pku.edu.cn/index.php/publications/frb180301/})

\section*{SUPPLEMENTARY DATA}
Supplementary data are available below and at \href{https://academic.oup.com/nsr/article/12/2/nwae293/7754190#supplementary-data}{NSR} online.

\section*{ACKNOWLEDGEMENTS}
This work made use of the data from FAST (Five-hundred-meter Aperture Spherical radio Telescope) (\url{https://cstr.cn/31116.02.FAST}). FAST is a Chinese national mega-science facility, operated by National Astronomical Observatories, Chinese Academy of Sciences.

\section*{FUNDING}
This work is supported by National SKA Program of China (2020SKA0120100,
2020SKA0120200), Natural Science Foundation of China (12041304, 11873067,
11988101, 12041303, 11725313, 11725314, 11833003, 12003028, 12041306,
12103089, U2031209, U2038105, U1831207), KJL acknowledges  funding from
the Max-Planck Partner Group and support from the XPLORER PRIZE.

\section*{AUTHOR CONTRIBUTIONS} JCJ and JWX led the data analysis of
polarization and drafted the manuscript, JRN worked on the PA jump, KJL,
WWZ, and BZ coordinated the observational campaign, cosupervised data
analyses and interpretations, and led the paper writing.  HX, DJZ, YKZ,
BJW, SC and CFZ contributed to radio data analysis. YPM contributed to
the searching software development. HQG, JLH, HL, LQ, JHS, RY, YLY, DJY,
and YZ aided with FAST observations and coordination. KJL, BZ, YQ, SSC,
WYW, DZL, RXX, WL, YPY, YL, and RL provided theoretical discussions. LFH,
YXH, ZXL, and YHX tested searching software.

\paragraph{\it Conflict of interest statement} None declared.

\clearpage
\onecolumn
\setlength{\oddsidemargin}{0.5cm}
\setcounter{figure}{0}
\setcounter{table}{0}
\setcounter{equation}{0}
\setcounter{section}{0}
\renewcommand*{\thefigure}{S\arabic{figure}}
\renewcommand*{\thetable}{S\arabic{table}}
\renewcommand*{\theequation}{S\arabic{equation}}
\renewcommand*{\thesection}{S\arabic{section}}
{\titlefont\centering Supplementary data for\\ Ninety percent circular polarization detected in a repeating fast radio burst \par}

\section{Radio observations and burst detection}\label{sec:s1}

FAST observed FRB~20201124A between September 25 and October 17, 
2021\cite{2022RAA....22l4001Z,2022RAA....22l4002Z,2022RAA....22l4003J,2022RAA....22l4004N}.  
In total, we have 19 hours of observations. Bursts were only detected between 
September 25 and 28. We pointed the central beam of the 19-beam receiver of 
FAST\cite{FASTperformance} to the coordinate provided by EVN, 
$\mathrm{RA}=05^\mathrm{h}08^\mathrm{m}03.5077^\mathrm{s}$, 
$\mathrm{Dec}=+26^\circ 03' 38.504''$\cite{EVN2021ATel14603}. The 19-beam 
receiver covers the frequency range of 1 -- 1.5~GHz. Search-mode filterbank data 
with four Stokes parameters were recorded for polarimetry studies using the 
pulsar digital backend\cite{FASTcommission,FASTperformance} in \texttt{PSRFITS} 
format\cite{PSRCHIVE2004PASA}. The full band, i.e. 1 to 1.5 GHz, was divided 
into 4096 frequency channels, and we record data at the temporal resolution of 
49.152 $\mu$s per sample.

We follow the rather standard procedures\cite{Luo2020Natur, 2022Natur.609..685X} 
to search for burst signals using the software package 
\texttt{TransientX}\cite{Men2019MNRAS,2024A&A...683A.183M}. The software can be 
found in the `Code Availability' section. The dispersion delay in cold plasma of 
free electrons is,
\begin{equation}
\Delta t=\mathcal{D}\frac{\mathrm{DM}}{\nu^2},
\end{equation}
where $\mathrm{DM}$ is dispersion measure, $\nu$ is frequency, and constant 
$\mathcal D= 4.148808\times 10^3\,\mathrm{MHz^2\,pc^{-1}\,cm^3\,s}$.  Pulses 
were de-dispersed with the daily average DM value published 
previously\cite{2022RAA....22l4001Z}, which are 412.4, 412.2, 412.5 and 
411.6~$\mathrm{pc\,cm^{-3}}$ for the 4 days from the 25th to the 28th September, 
respectively. Then, the burst signals were searched using boxcar matched filters 
with widths ranging from 0.1 to 100 ms with the signal-to-noise ratio threshold 
of 7. RFIs were mitigated in the search. The continuous narrow band satellite 
RFIs are removed according to the RFI list, and a zero-DM matched filter is used 
to identify wideband RFIs without dispersion\cite{Men2019MNRAS}. The searching 
results were later verified by human inspection so that the collected bursts are 
with a clear dispersive signature. Once the bursts were detected, we further 
removed data in the 20 MHz band edges at both lower and upper frequency ends for 
our later analysis.

\section{Polarization calibration}\label{sec:s2}

The 19-beam receiver of FAST uses orthogonal dual linear polarization feeds to 
receive radio signals\cite{FASTperformance}. In the first and the last minutes 
of the observation, modulated signals from a noise diode were injected as a 
$45^\circ$ linearly polarized calibrator. The period of modulated noise signals 
was 100.663296 ms and their duty cycle was 50\%. We used software package 
\texttt{DSPSR}\cite{DSPSR2011PASA} to fold the modulated noise signals, and then 
used software package \texttt{PSRCHIVE}\cite{PSRCHIVE2004PASA,PSRCHIVE2012AR&T} 
to calibrate polarization data with the single-axial model. After the 
polarimetric calibration, the systematical error is at the level of 
0.5\%\cite{Dunning2017, Luo2020Natur, 2022Natur.609..685X}. In this manuscript, 
we adopt the PSR/IEEE convention\cite{PSRCHIVE2010PASA} for the definitions of 
the Stokes parameters. We checked and found that the polarization properties are essentially not affected by the DM values we use. For most bursts, the polarization profiles using daily averaged DM or DM from maximizing the structure show no visual differences. Most of the bursts that show large differences are due to their complex structures in the dynamic spectra, for which DM inference is unstable. In this paper, we show the polarisation measurements with daily averaged DM.

\section{Measurement and Correction of Faraday Rotation}\label{sec:s3}

Linear polarization is rotated with frequency, when radio waves propagate 
through a magnetized plasma, The change of the position angle is,
\begin{equation}
\Delta{\rm PA}=\mathrm{RM}\,\lambda^2,
\end{equation}
where $\lambda$ is the wavelength of the radio wave. Similar to previous 
work\cite{2022Natur.609..685X}, we used the Bayesian 
method\cite{Desvignes2019Sci,Luo2020Natur} to fit the rotation of Stokes $Q$ and 
$U$ as a function of frequencies to derive the rotation measures (RMs) of the 
bursts. We also cross checked the results with the revisited rotation measure 
synthesis method\cite{Schnitzeler2015MNRAS}. We had corrected the RM 
contribution caused by the geomagnetic field and Earth ionosphere using the 
estimation from the software package \texttt{ionFR}\cite{ionfr}. Linearly 
polarized intensity $L$, polarized intensity $P$ and PA were calculated after 
correcting Faraday rotation to the infinite frequency. The generalized Weisberg 
correction \cite{Everett2001ApJ,2022RAA....22l4003J} was applied when calculating $L$ and $P$. 

\section{Flux Density and Burst Energy}\label{sec:s4}
Our flux calibration for all the bursts is 
identical with the previous work. For the details of flux, 
we refer readers to Ref.\cite{2020Natur.582..351C, 2022Natur.609..685X, 2022RAA....22l4003J}. The flux densities $S$ were 
estimated using the radiometer equation,
\begin{equation}
    S_\nu=\frac{T_\mathrm{sys}(\mathrm{S/N})}{G\sqrt{2B\tau}},
\end{equation}
where the system temperature $T_\mathrm{sys}\simeq 24\,\mathrm{K}$, gain 
$G\approx 16\,\mathrm{K\,Jy^{-1}}$\cite{FASTperformance}, and $B$ and $\tau$ are 
the signal bandwidth and burst duration, respectively.
The fluences $F$ are calculated by integrating the flux densities over the pulse 
duration. We compute the burst fluence using three approaches: 1) we average the detected signal within the 500~MHz observing band and 2) we fit the spectrum of a burst with a Gaussian function, then we estimate the fluence using the fitted Gaussian function. Both methods can be affected by scintillation and the limited observing bandwidth, as the true signal central frequency may lay outside of the observing band. In order to avoid the problem, we also take the third approach by generating a subset of the data with the central frequency inside the observing band. However, the number of bursts in the sample is limited. The distribution of burst fluence is shown in \FIGU{fig:edocp}, where we also compare the fluence distributions for the 
bursts with high and low degrees of circular polarization. We note that the conclusion is not affected by how we compute the fluence.

To check if the fluence distribution of the bursts with high degree circular 
polarization ($\Pi_{\rm v}> 50\%$) are different comparing to that of the bursts 
with $\Pi_{\rm v}\le 50\%$,
we perform the standard two-sample Kolmogorov-Smirnov test\cite{NR3} between the 
fluence distributions. Here, the null hypothesis claims that the fluence 
distributions of large and small degrees of circular polarization are identical, 
while the alternative claims that the fluence distributions are different. We get 
a p-value of 0.39 for fluence averaged in the observing band, 0.95 for the the Gaussian fitted spectral peak fluence for all the bursts, and 0.30 for the Gaussian fitted spectral peak fluence for the bursts whose peak frequency falls in the observing band in the Kolmogorov-Smirnov test, which cannot reject the null 
hypothesis, i.e. we cannot differentiate the fluence distributions of the two samples. 

To investigate if the null result is caused by the limited number of samples, we 
performed simulations to evaluate the statistical power of the 
Kolmogorov-Smirnov test for the current limited sample of bursts. In our 
simulation, an fluence factor ($\le 1$) is multiplied to the fluence of high 
$\Pi_{\rm v}$ bursts, and then we perform the Kolmogorov-Smirnov test and 
compute the p-value. The procedure is carried out for a uniform grid of fluence 
factor from 0.01 to 1.0.  We can thus compute the fluence factors at which the 
null hypothesis can be rejected for a given confidence level. As shown in 
\FIGU{fig:kssens}, one would claim a 2-sigma detection, i.e. p-value of 0.05, 
when the energy of high $\Pi_{\rm v}$ burst drop to approximately 65\% for fluence averaged in the observing bnad, 35\% for the Gaussian fitted fluence peak for all the bursts, and 8\% for the Gaussian fitted fluence peak for the bursts whose peak falls in the observing band. In this 
way, our data set is sensitive to 35\% fluence difference between the  high and 
low $\Pi_{\rm v}$ bursts, i.e. we would detect the fluence difference of the two 
population,  if the high $\Pi_{\rm v}$ bursts were 35\% weaker. Thus, any 
theoretical model predicting significant correlation between burst energy and 
$\Pi_{\rm v}$ would be in tension with our observations.

\section{Astrophysical sources with significant circular polarization}\label{sec:s5}

In Figure~2, we collected the brightness temperature and degree of circular 
polarization of various astrophysical sources, including solar radio bursts, 
Jupiter, radio pulsars, and FRBs. Our data source for solar radio bursts are 
from Ref.\cite{Dulk1985ARA&A}. For radio emission from Jupiter and its 
magnetosphere, there are three possibilities\cite{2016A&A...587A...3G}: i) 
thermal emission from planetary disks, ii) auroral cyclotron emission below 
40~MHz, including decameter (DAM), hectometer (HOM) and broadband kilometer 
(bKOM) emissions, iii) synchrotron emission between 30~MHz and 30~GHz. Because 
we focus on the burst-like sources, we only include the Jovian DAM emission in 
Figure~2.  The brightness temperature of Jovian DAM can reach 
$>10^{15}\,\mathrm{K}$\cite{2014JGRA..119.9508C}, while the degrees of circular 
polarization of some bursts approach 100\%\cite{2017A&A...604A..17M}.

A fraction of pulsars exhibit strong circular polarization\cite{Han1998MNRAS}.  
We include the polarization observation of normal 
pulsars\cite{2018MNRAS.474.4629J} and millisecond 
pulsars\cite{2022PASA...39...27S} in Figure~2. We collected the period, flux and 
pulse width from the ATNF Pulsar Catalogue\cite{psrcat}. The brightness 
temperature of a pulsar is estimated\cite{handbook} using  
$T_b=SD^2c^2/2kA\nu^2$, where $S$ is the observed radio flux, $D$ is the pulsar 
distance. $A$ is the effective magnetosphere area and estimated using the pulse 
timescale, i.e. $A=\pi W^2c^2$, with $W$ being the pulse width at half of peak.  
Constants $k$ and $c$ are the Boltzmann constant and light speed, respectively.

For FRBs, non-detection or marginal detection of circular polarization has been 
previously reported for FRB repeaters, e.g. 
FRB~20121102A\cite{Michilli2018Natur,Gajjar2018ApJ,2021RNAAS...5...17F,Hilmarsson2021ApJ},  
FRB~20180301A\cite{2023MNRAS.526.3652K}, 20190711A\cite{Day2020MNRAS}, and 
20220912A\cite{2023MNRAS.526.2039H}.  In addition to FRB~20201124A, circularly 
polarized emission has been sporadically detected from a few FRB repeaters. 
Circular polarization has also been reported in bursts from 
FRB~20220912A\cite{2023ApJ...955..142Z,2024ApJ...974..296F} (70\%) and 
FRB~20190520B\cite{2023Sci...380..599A} ($42\%$).  Recently, circular 
polarization of $64\%$ from FRB~20121102A was also 
reported\cite{2022SciBu..67.2398F}.  In Figure~2, the highest values of 
published $\Pi_{\rm v}$ are collected. The brightness temperature of FRB is 
estimated\cite{Zhang2023RMP} using \begin{equation}
    T_b=\frac{SD_\mathrm{L}^2}{2\pi k_\mathrm{B}(1+z)\nu^2\Delta t^2},\label{eq:brightness_temperature}
\end{equation}
where $D_\mathrm{L}$ is the luminosity distance derived from the redshift $z$, 
and, similar to the pulsar case, the burst time width $\Delta t$ is used to 
estimate the size of emission zone\cite{Zhang2023RMP}.

\clearpage

\begin{figure}
    \centering
		\includegraphics[width=170mm]{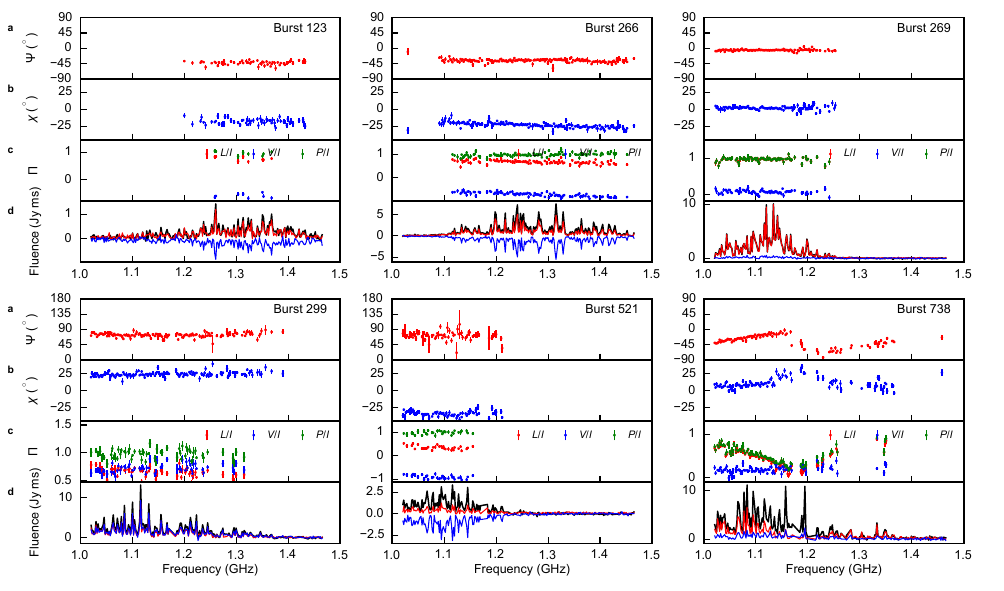}
		\caption{polarization spectra of selected sample of bursts with high degrees 
		of circular polarization or abrupt jumps in linear polarization position 
		angle. For each burst, we plotted (\textbf{a}) Position angle, (\textbf{b}) 
		Ellipticity angle, (\textbf{c}) degree of linear (red), circular (blue) and 
		total polarization (green), and (\textbf{d}) Burst fluence of total intensity 
		(black), linearly polarization (red), and circular polarization (blue) as 
		functions of frequency.
		}
\label{fig:spectral}
\end{figure}

\begin{figure}
\centering
\includegraphics[width=170mm]{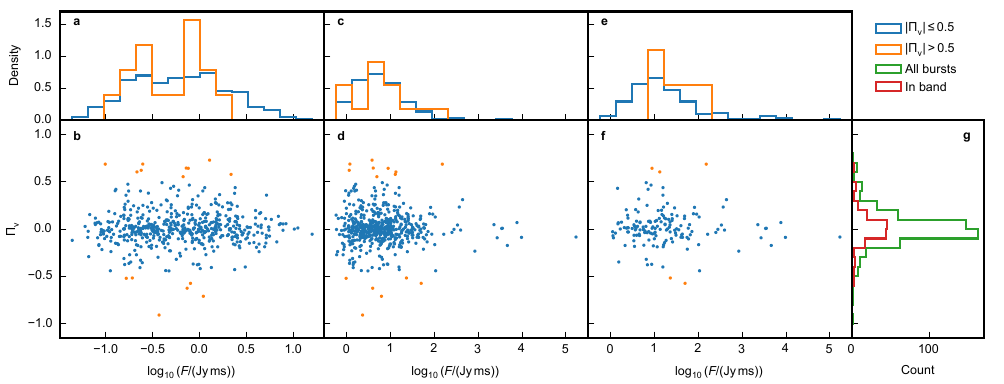}
\caption{The distribution of burst fluence and degree of circular polarization 
($\Pi_{\rm v}$) for bursts with $S/N\ge 50$. (\textbf{a}) histogram of burst fluence averaged in the observing band, 
bursts with $|\Pi_{\rm v}|\ge 50\%$ are in orange, while the other bursts are in 
blue. (\textbf{b}) two dimensional distribution of burst fluence averaged in the observing band and $\Pi_{\rm v}$ with the same 
color convention of a). (\textbf{c}) histogram of burst fluence using Gaussian fitted spectra for all the bursts. (\textbf{d}) two dimensional distribution of fluence using Gaussian fitted spectra and $\Pi_{\rm v}$ for all the bursts. (\textbf{e}) histogram of burst fluence using Gaussian fitted spectra for the bursts with spectral peak in the observing band. (\textbf{f}) two dimensional distribution of fluence using Gaussian fitted spectra and $\Pi_{\rm v}$ for the bursts with spectral peak in the observing band. (\textbf{g}) Histogram of $\Pi_{\rm v}$ for all bursts and bursts with spectral peak in the observing band.
}\label{fig:edocp}
\end{figure}

\begin{figure}
\centering
\includegraphics[width=65mm]{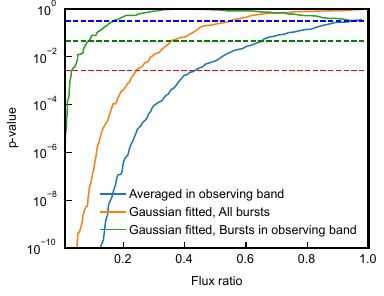}
\caption{The relation between the p-value of Kolmogorov-Smirnov test and fluence 
factor multiplied to the fluence of bursts with $|\Pi_{\rm v}|\ge 50\%$. The blue, orange and green solid 
curves are the p-value and fluence factor relation for burst fluence averaged in the observing band, Gaussian fitted spectra for all bursts and bursts with spectral peaks in the observing band, respectively. The blue, green, and red 
dashed lines are for 1-, 2-, and 3-$\sigma$, i.e. p-value of 32\%, 5\%, and 0.3 
\%, respectively.
}\label{fig:kssens}
\end{figure}

\begin{figure}
    \centering
		\includegraphics[width=170mm]{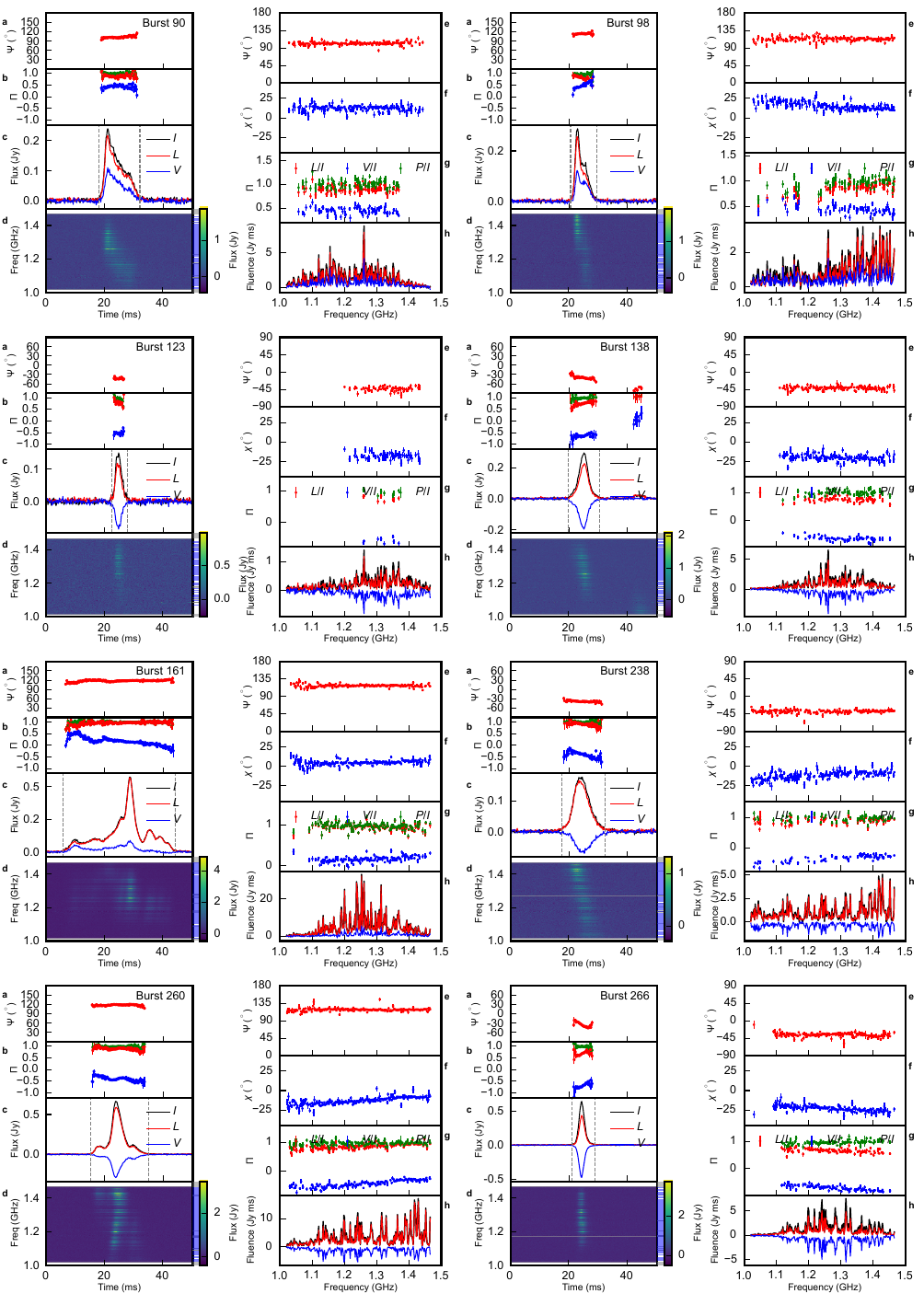}
		\caption{Bursts with $|\Pi_{\rm V}| \ge 50\%$ and S/N$\ge 50$. Notations are 
		the same as in Figure~1.
		}\label{fig:highvatlas1}
\end{figure}
\begin{figure}
\includegraphics[width=170mm]{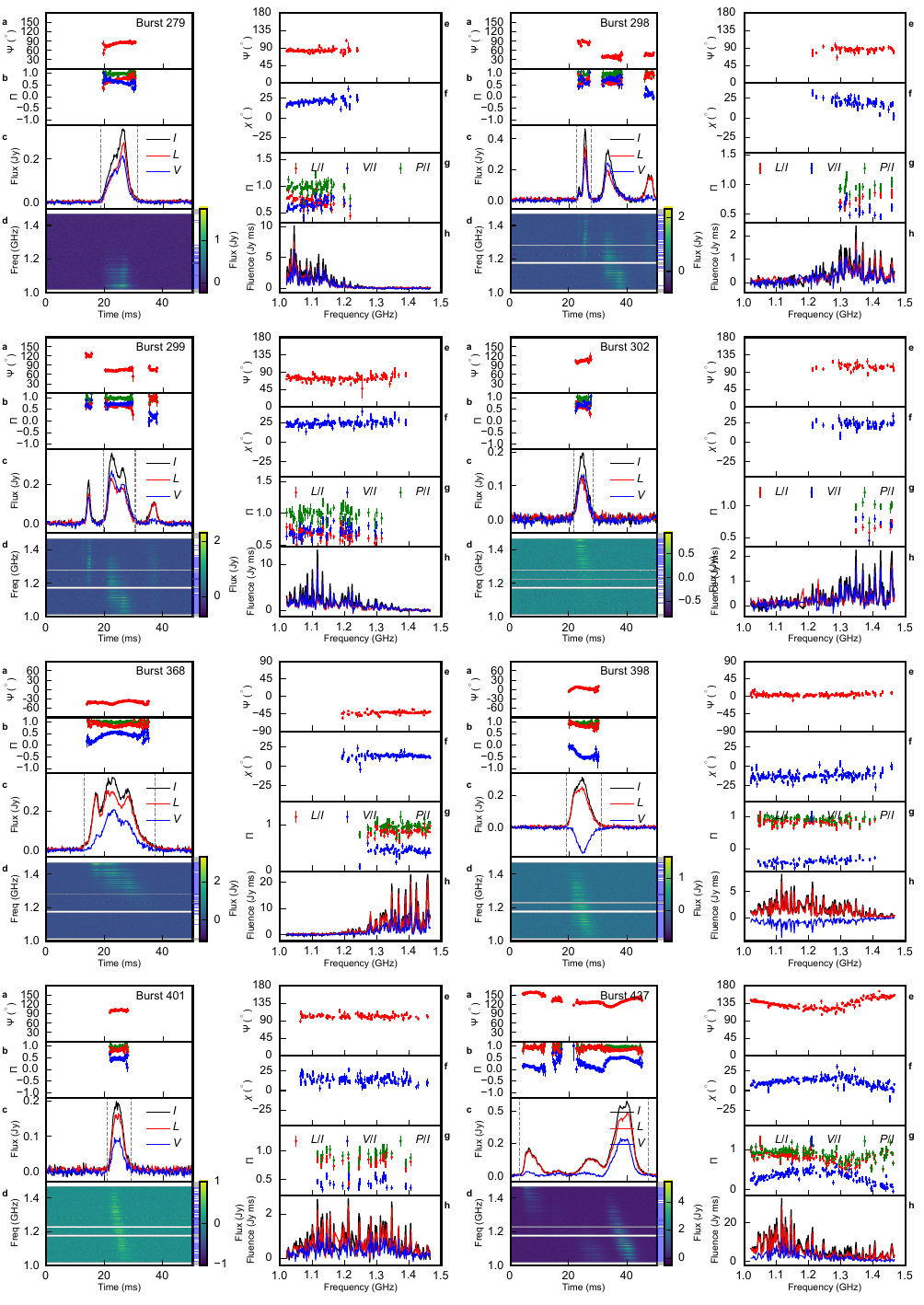}
		\caption{Continuation of \FIGU{fig:highvatlas1}.
		}\label{fig:highvatlas2}
\end{figure}
\begin{figure}
    \centering
\includegraphics[width=170mm]{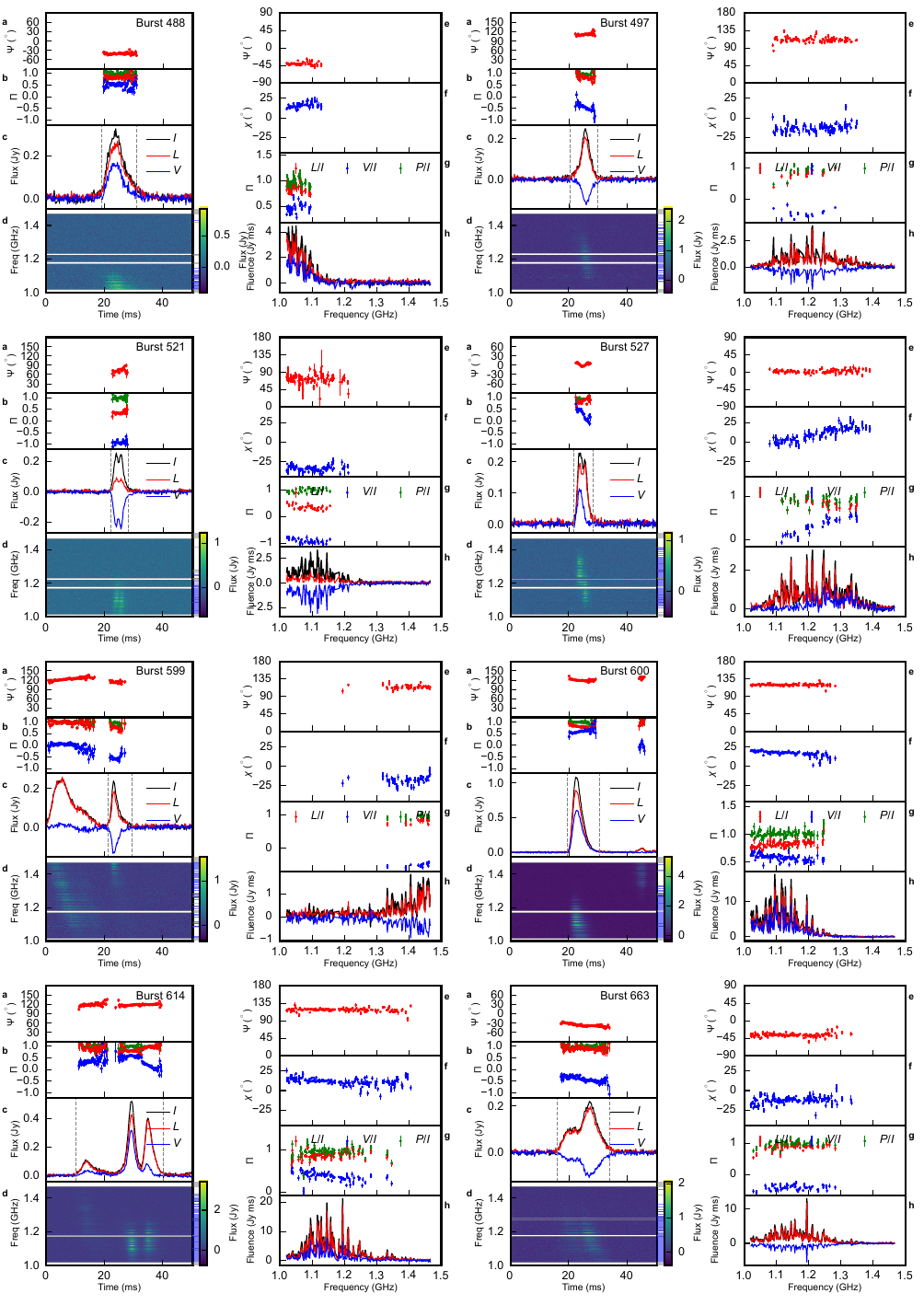}
		\caption{Continuation of \FIGU{fig:highvatlas1}.
		}\label{fig:highvatlas3}
\end{figure}
\begin{figure}
\centering
\includegraphics[width=170mm]{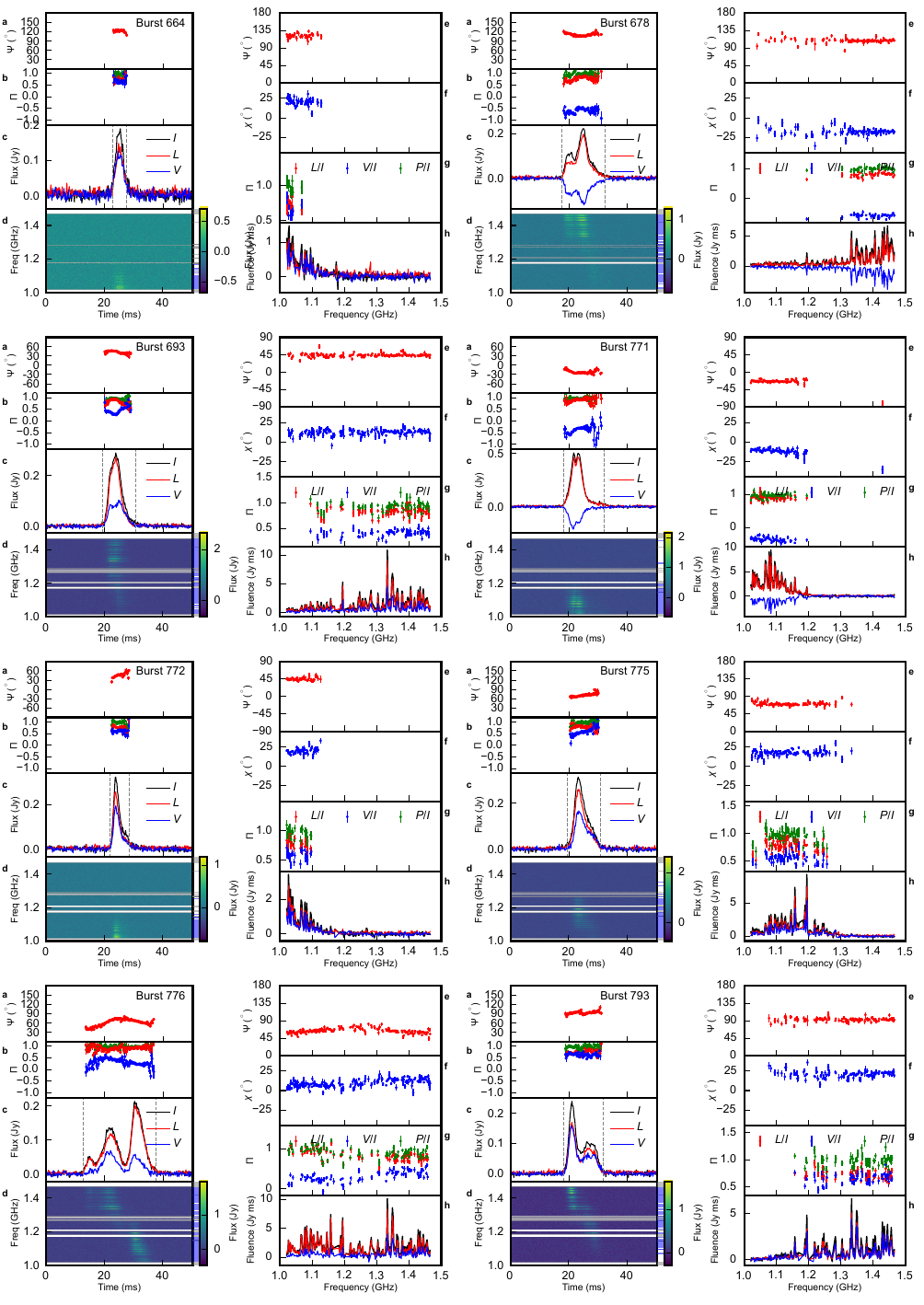}
\caption{Continuation of \FIGU{fig:highvatlas1}.  }
\label{fig:highvatlas4}
\end{figure}
\clearpage

\begin{adjustbox}{center}
\begin{threeparttable}
\scriptsize

\caption{Summary of high $\Pi_{\rm v}$ burst properties. Uncertainties are for 
68\% confidence interval.}
 \label{tab:summary}
\begin{tabular}{c c c c c c c c c }
\hline
\hline
No.\tnote{a} &TOA\tnote{b} & $\mathrm{RM_{Bayes}}$\tnote{c} & $(L/I)$\tnote{d} & $(V/I)_\mathrm{mean}$\tnote{e} & $(V/I)_\mathrm{peak}$\tnote{f}  & $P/I$\tnote{g}& Peak flux& Fluence\tnote{h}\\
 &day & $\mathrm{rad\,m^{-2}}$  & \% & \% & \% & \% & Jy & Jy ms\\
\hline
90 & 0.818177  & $-584.9 \pm0.5$ & $88.3 \pm0.7$ & $43.7 \pm0.5$ & $56.5 \pm3.5$ & $98.5 \pm0.7$ & 0.51 & 1.5\\
98 & 0.819599 & $-591.7 \pm0.5$ & $82.2 \pm0.7$ & $46.5 \pm0.6$ & $61 \pm5$ & $94.4 \pm0.8$ & 0.66 & 1.0\\
123 & 0.825277 & $-578.9 \pm2.3$ & $75.6 \pm1.7$ & $-52.1 \pm1.6$ & $-57 \pm5$ & $91.8 \pm1.9$ & 0.35 & 0.4\\
138 & 0.828698  & $-583.5 \pm0.8$ & $75.0 \pm0.7$ & $-62.6 \pm0.6$ & $-68.8 \pm2.3$ & $97.7 \pm0.7$ & 0.69 & 1.2\\
161 & 0.833527  & $-594.23 \pm0.28$ & $96.34 \pm0.26$ & $16.38 \pm0.19$ & $58 \pm5$ & $97.72 \pm0.26$ & 1.40 & 5.3\\
238 & 0.845965  & $-590.38 \pm0.35$& $91.4 \pm0.7$ & $-37.8 \pm0.5$ & $-56 \pm4$ & $99.0 \pm0.7$ & 0.47 & 1.2\\
260 & 0.849867  & $-593.39 \pm0.2$ & $89.50 \pm0.24$ & $-40.12 \pm0.19$ & $-58.3 \pm3.5$ & $98.08 \pm0.25$ & 1.76 & 3.5\\
266 & 0.851199  & $-595.6 \pm0.6$  & $66.86 \pm0.35$ & $-71.1 \pm0.4$ & $-80.3 \pm3.5$ & $97.6 \pm0.4$ & 1.70 & 1.5\\
279 & 0.853596 & $-594.2 \pm0.7$  & $73.0 \pm0.5$ & $64.2 \pm0.5$ & $83 \pm4$ & $97.3 \pm0.6$ & 0.90 & 1.9\\
298 & 1.786479  & $-585.3 \,^{+2.7}_{-3.0}$ & $71.2 \pm1.5$ & $62.1 \pm1.4$ & $72.6 \pm3.2$ & $94.5 \pm1.7$ & 0.73 & 0.7\\
299 & 1.786479  & $-596.2 \pm0.7$  & $65.6 \pm0.6$ & $72.9 \pm0.6$ & $78.8 \pm2.9$ & $98.0 \pm0.7$ & 0.50 & 2.1\\
302 & 1.786862  & $-612.6 \pm3.4$ & $61.1 \pm1.6$ & $68.7 \pm1.7$ & -- & $91.9 \pm1.8$ & 0.32 & 0.7\\
368 & 1.792096  & $-622.1 \pm0.7$ & $87.0 \pm0.4$ & $43.79 \pm0.34$ & $59.7 \pm3.3$ & $97.4 \pm0.4$ & 0.66 & 4.8\\
398 & 1.794975  & $-613.0 \pm0.4$  & $84.3 \pm0.6$ & $-40.7 \pm0.5$ & $-58.1 \pm2.6$ & $93.6 \pm0.7$ & 0.48 & 1.8\\
401 & 1.795033  & $-621.0 \pm0.8$  & $83.3 \pm1.2$ & $45.2 \pm1.0$ & $54 \pm4$ & $94.8 \pm1.3$ & 0.32 & 0.9\\
437 & 1.797915  & $-637.9 \pm1.2$  & $79.52 \pm0.32$ & $35.83 \pm0.27$ & $54.9 \pm1.2$ & $87.21 \pm0.33$ & 0.85 & 6.2\\
488 & 1.802788  & $-588.4 \,^{+2.1}_{-2.0}$ & $82.4 \pm1.1$ & $49.0 \pm1.0$ & $59 \pm5$ & $95.9 \pm1.2$ & 3.62 & 2.0\\
497 & 1.803396  & $-579.7 \pm1.3$  & $81.8 \pm1.3$ & $-43.9 \pm1.1$ & $-55 \pm4$ & $92.9 \pm1.4$ & 0.42 & 0.9\\
521 & 1.804982  & $-568.4 \,^{+2.8}_{-2.7}$  & $34.3 \pm0.9$ & $-90.9 \pm1.1$ & $-100 \pm4$ & $97.2 \pm1.2$ & 0.52 & 1.1\\
527 & 1.805791  & $-615.4 \pm0.9$  & $84.3 \pm1.1$ & $30.7 \pm0.9$ & $52.4 \pm3.3$ & $89.7 \pm1.1$ & 0.41 & 0.8\\
599 & 1.810067  & $-610.8 \,^{+3.3}_{-3.5}$  & $74.0 \pm1.7$ & $-51.7 \pm1.6$ & $-61 \pm4$ & $90.3 \pm1.9$ & 0.51 & 0.7\\
600 & 1.810074  & $-613.0 \pm0.4$  & $81.07 \pm0.28$ & $57.80 \pm0.25$ & $69.8 \pm2.4$ & $99.56 \pm0.30$ & 1.90 & 4.2\\
614 & 1.810844  & $-619.3 \pm0.5$  & $87.7 \pm0.5$ & $37.8 \pm0.4$ & $62.2 \pm1.3$ & $95.5 \pm0.5$ & 0.92 & 3.6\\
663 & 1.814971  & $-595.3 \pm0.7$ & $88.3 \pm0.7$ & $-41.8 \pm0.6$ & $-55 \pm4$ & $97.7 \pm0.8$ & 0.42 & 1.9\\
664 & 1.815084  & $-612.2 \,^{+3.4}_{-3.5}$ & $74.3 \pm2.2$ & $68.6 \pm2.1$ & -- & $101.1 \pm2.5$ & 0.36 & 0.6\\
678 & 1.815578 & $-614.2 \pm1.0$  & $78.2 \pm0.8$ & $-57.4 \pm0.7$ & $-59.8 \pm3.2$ & $97.0 \pm0.9$ & 0.50 & 1.3\\
693 & 1.817261 & $-594.9 \pm0.4$  & $83.6 \pm0.6$ & $40.3 \pm0.5$ & $58.4 \pm3.3$ & $92.8 \pm0.6$ & 0.64 & 1.4\\
771 & 1.823548  & $-597.3 \pm0.8$  & $89.6 \pm0.6$ & $-37.0 \pm0.4$ & $-60.5 \pm3.1$ & $97.0 \pm0.6$ & 1.08 & 2.5\\
772 & 1.823569  & $-583.8 \,^{+1.8}_{-1.7}$  & $75.4 \pm1.2$ & $60.4 \pm1.1$ & $66 \pm4$ & $96.6 \pm1.3$ & 0.70 & 0.9\\
775 & 1.823892  & $-604.8 \pm0.7$  & $78.4 \pm0.7$ & $55.3 \pm0.6$ & $62 \pm4$ & $96.0 \pm0.7$ & 0.67 & 1.5\\
776 & 1.824116  & $-613.0 \pm0.5$  & $87.6 \pm0.6$ & $31.7 \pm0.5$ & $54 \pm4$ & $93.2 \pm0.6$ & 0.46 & 2.0\\
793 & 1.824974  & $-594.8 \pm0.7$  & $71.9 \pm0.7$ & $64.5 \pm0.7$ & $74 \pm4$ & $96.6 \pm0.8$ & 0.60 & 1.3\\
\hline
\hline
\end{tabular}
\begin{tablenotes}
\item[a] Burst index number.
\item[b] Barycentric burst time of arrival in Barycentric Coordinate Time (TCB) scale. The time reference of $T=0$ is MJD 59484.  
\item[c] Faraday rotation measure using Bayesian method after correcting the Earth ionosphere contribution.
\item[d] Average degree of linear polarization.
\item[e] Average degree of circular polarization.
\item[f] Peak degree of circular polarization.
\item[g] Average degree of total polarization.
\item[h] Average fluence in the burst frequency channels.
\end{tablenotes}
\end{threeparttable}
\end{adjustbox}
\bibliographystyle{nsr}
\bibliography{ms}

\begin{thebibliography}{10}

\bibitem{2020Natur.587...59B}
{Bochenek} CD, {Ravi} V, {Belov} KV {\em et~al.}
\newblock {A fast radio burst associated with a Galactic magnetar}.
\newblock {\em \nat}  2020; {\bf 587}: 59--62.

\bibitem{2020Natur.587...54C}
{CHIME/FRB Collaboration}, {Andersen} BC, {Bandura} KM {\em et~al.}
\newblock {A bright millisecond-duration radio burst from a Galactic magnetar}.
\newblock {\em \nat}  2020; {\bf 587}: 54--58.

\bibitem{Zhang2020Natur}
{Zhang} B.
\newblock {The physical mechanisms of fast radio bursts}.
\newblock {\em \nat}  2020; {\bf 587}: 45--53.

\bibitem{Metzger19}
{Metzger} BD, {Margalit} B and {Sironi} L.
\newblock {Fast radio bursts as synchrotron maser emission from decelerating
  relativistic blast waves}.
\newblock {\em \mnras}  2019; {\bf 485}: 4091--4106.

\bibitem{Beloborodov20}
{Beloborodov} AM.
\newblock {Blast Waves from Magnetar Flares and Fast Radio Bursts}.
\newblock {\em \apjs}  2020; {\bf 896}:142.

\bibitem{Kumar17}
{Kumar} P, {Lu} W and {Bhattacharya} M.
\newblock {Fast radio burst source properties and curvature radiation model}.
\newblock {\em \mnras}  2017; {\bf 468}: 2726--2739.

\bibitem{Yang18}
{Yang} YP and {Zhang} B.
\newblock {Bunching Coherent Curvature Radiation in Three-dimensional Magnetic
  Field Geometry: Application to Pulsars and Fast Radio Bursts}.
\newblock {\em \apjs}  2018; {\bf 868}:31.

\bibitem{2022ApJ...927..105W}
{Wang} WY, {Yang} YP, {Niu} CH {\em et~al.}
\newblock {Magnetospheric Curvature Radiation by Bunches as Emission Mechanism
  for Repeating Fast Radio Bursts}.
\newblock {\em \apj}  2022; {\bf 927}:105.

\bibitem{2022SCPMA..6589511W}
{Wang} WY, {Jiang} JC, {Lu} J {\em et~al.}
\newblock {Repeating fast radio bursts: Coherent circular polarization by
  bunches}.
\newblock {\em Science China Physics, Mechanics, and Astronomy}  2022; {\bf
  65}:289511.

\bibitem{2022ApJ...925...53Z}
{Zhang} B.
\newblock {Coherent Inverse Compton Scattering by Bunches in Fast Radio
  Bursts}.
\newblock {\em \apj}  2022; {\bf 925}:53.

\bibitem{2023MNRAS.522.2448Q}
{Qu} Y and {Zhang} B.
\newblock {Polarization of fast radio bursts: radiation mechanisms and
  propagation effects}.
\newblock {\em \mnras}  2023; {\bf 522}: 2448--2477.

\bibitem{2019MNRAS.486.3636P}
{Price} DC, {Foster} G, {Geyer} M {\em et~al.}
\newblock {A fast radio burst with frequency-dependent polarization detected
  during Breakthrough Listen observations}.
\newblock {\em \mnras}  2019; {\bf 486}: 3636--3646.

\bibitem{Luo2020Natur}
{Luo} R, {Wang} BJ, {Men} YP {\em et~al.}
\newblock {Diverse polarization angle swings from a repeating fast radio burst
  source}.
\newblock {\em \nat}  2020; {\bf 586}: 693--696.

\bibitem{2022Natur.609..685X}
{Xu} H, {Niu} JR, {Chen} P {\em et~al.}
\newblock {A fast radio burst source at a complex magnetized site in a barred
  galaxy}.
\newblock {\em \nat}  2022; {\bf 609}: 685--688.

\bibitem{Hilmarsson2021MNRAS}
{Hilmarsson} GH, {Spitler} LG, {Main} RA {\em et~al.}
\newblock {Polarization properties of FRB 20201124A from detections with the
  Effelsberg 100-m radio telescope}.
\newblock {\em \mnras}  2021; {\bf 508}: 5354--5361.

\bibitem{Kumar2022MNRAS}
{Kumar} P, {Shannon} RM, {Lower} ME {\em et~al.}
\newblock {Circularly polarized radio emission from the repeating fast radio
  burst source FRB 20201124A}.
\newblock {\em \mnras}  2022; {\bf 512}: 3400--3413.

\bibitem{2022RAA....22l4003J}
{Jiang} JC, {Wang} WY, {Xu} H {\em et~al.}
\newblock {FAST Observations of an Extremely Active Episode of FRB 20201124A.
  III. Polarimetry}.
\newblock {\em Research in Astronomy and Astrophysics}  2022; {\bf 22}:124003.

\bibitem{2023PhRvD.108d3009K}
{Kumar} P, {Shannon} RM, {Lower} ME {\em et~al.}
\newblock {Propagation of a fast radio burst through a birefringent
  relativistic plasma}.
\newblock {\em \prd}  2023; {\bf 108}:043009.

\bibitem{2024ApJ...974..296F}
{Feng} Y, {Li} D, {Zhang} YK {\em et~al.}
\newblock {An Extremely Active Repeating Fast Radio Burst Source in a Likely
  Nonmagneto-ionic Environment}.
\newblock {\em \apj}  2024; {\bf 974}:296.

\bibitem{Masui2015Natur}
{Masui} K, {Lin} HH, {Sievers} J {\em et~al.}
\newblock {Dense magnetized plasma associated with a fast radio burst}.
\newblock {\em \nat}  2015; {\bf 528}: 523--525.

\bibitem{Petroff2015MNRAS}
{Petroff} E, {Bailes} M, {Barr} ED {\em et~al.}
\newblock {A real-time fast radio burst: polarization detection and
  multiwavelength follow-up}.
\newblock {\em \mnras}  2015; {\bf 447}: 246--255.

\bibitem{Caleb2018MNRAS}
{Caleb} M, {Keane} EF, {van Straten} W {\em et~al.}
\newblock {The SUrvey for Pulsars and Extragalactic Radio Bursts - III.
  Polarization properties of FRBs 160102 and 151230}.
\newblock {\em \mnras}  2018; {\bf 478}: 2046--2055.

\bibitem{Oslowski2019MNRAS}
{Os{\l}owski} S, {Shannon} RM, {Ravi} V {\em et~al.}
\newblock {Commensal discovery of four fast radio bursts during Parkes Pulsar
  Timing Array observations}.
\newblock {\em \mnras}  2019; {\bf 488}: 868--875.

\bibitem{Day2020MNRAS}
{Day} CK, {Deller} AT, {Shannon} RM {\em et~al.}
\newblock {High time resolution and polarization properties of ASKAP-localized
  fast radio bursts}.
\newblock {\em \mnras}  2020; {\bf 497}: 3335--3350.

\bibitem{Prochaska2019Sci}
{Prochaska} JX, {Macquart} JP, {McQuinn} M {\em et~al.}
\newblock {The low density and magnetization of a massive galaxy halo exposed
  by a fast radio burst}.
\newblock {\em Science}  2019; {\bf 366}: 231--234.

\bibitem{Cho2020ApJ}
{Cho} H, {Macquart} JP, {Shannon} RM {\em et~al.}
\newblock {Spectropolarimetric Analysis of FRB 181112 at Microsecond
  Resolution: Implications for Fast Radio Burst Emission Mechanism}.
\newblock {\em \apjl}  2020; {\bf 891}:L38.

\bibitem{Mckinven2021ApJ}
{Mckinven} R, {Michilli} D, {Masui} K {\em et~al.}
\newblock {Polarization Pipeline for Fast Radio Bursts Detected by CHIME/FRB}.
\newblock {\em \apjs}  2021; {\bf 920}:138.

\bibitem{2023Sci...380..599A}
{Anna-Thomas} R, {Connor} L, {Dai} S {\em et~al.}
\newblock {Magnetic field reversal in the turbulent environment around a
  repeating fast radio burst}.
\newblock {\em Science}  2023; {\bf 380}: 599--603.

\bibitem{2022SciBu..67.2398F}
{Feng} Y, {Zhang} YK, {Li} D {\em et~al.}
\newblock {Circular polarization in two active repeating fast radio bursts}.
\newblock {\em Science Bulletin}  2022; {\bf 67}: 2398--2401.

\bibitem{2023ApJ...955..142Z}
{Zhang} YK, {Li} D, {Zhang} B {\em et~al.}
\newblock {FAST Observations of FRB 20220912A: Burst Properties and
  Polarization Characteristics}.
\newblock {\em \apjs}  2023; {\bf 955}:142.

\bibitem{2023ApJ...949L...3R}
{Ravi} V, {Catha} M, {Chen} G {\em et~al.}
\newblock {Deep Synoptic Array Science: Discovery of the Host Galaxy of FRB
  20220912A}.
\newblock {\em \apjl}  2023; {\bf 949}:L3.

\bibitem{2022RAA....22l4001Z}
{Zhou} DJ, {Han} JL, {Zhang} B {\em et~al.}
\newblock {FAST Observations of an Extremely Active Episode of FRB 20201124A:
  I. Burst Morphology}.
\newblock {\em Research in Astronomy and Astrophysics}  2022; {\bf 22}:124001.

\bibitem{2022RAA....22l4002Z}
{Zhang} YK, {Wang} P, {Feng} Y {\em et~al.}
\newblock {FAST Observations of an Extremely Active Episode of FRB 20201124A.
  II. Energy Distribution}.
\newblock {\em Research in Astronomy and Astrophysics}  2022; {\bf 22}:124002.

\bibitem{2022RAA....22l4004N}
{Niu} JR, {Zhu} WW, {Zhang} B {\em et~al.}
\newblock {FAST Observations of an Extremely Active Episode of FRB 20201124A.
  IV. Spin Period Search}.
\newblock {\em Research in Astronomy and Astrophysics}  2022; {\bf 22}:124004.

\bibitem{CHIME2021ApJS}
{Amiri} M, {Andersen} BC, {Bandura} K {\em et~al.}
\newblock {The First CHIME/FRB Fast Radio Burst Catalog}.
\newblock {\em \apjs}  2021; {\bf 257}:59.

\bibitem{1985SSRv...41..215W}
{Wu} CS.
\newblock {Kinetic cyclotron and synchrotron maser instabilities: Radio
  emission processes by direct amplification of radiation}.
\newblock {\em Space Sci. Rev.}  1985; {\bf 41}: 215--298.

\bibitem{Wang10}
{Wang} C, {Lai} D and {Han} J.
\newblock {Polarization changes of pulsars due to wave propagation through
  magnetospheres}.
\newblock {\em \mnras}  2010; {\bf 403}: 569--588.

\bibitem{Lu21}
{Lu} JG, {Wang} WY, {Peng} B {\em et~al.}
\newblock {Are pulsar giant pulses induced by re-emission of cyclotron
  resonance absorption?}
\newblock {\em Research in Astronomy and Astrophysics}  2021; {\bf 21}:029.

\bibitem{2022NatCo..13.4382W}
{Wang} FY, {Zhang} GQ, {Dai} ZG {\em et~al.}
\newblock {Repeating fast radio burst 20201124A originates from a magnetar/Be
  star binary}.
\newblock {\em Nature Communications}  2022; {\bf 13}:4382.

\bibitem{2022ApJ...933L...6L}
{Lyutikov} M.
\newblock {Faraday Conversion in Pair-symmetric Winds of Magnetars and Fast
  Radio Bursts}.
\newblock {\em \apjl}  2022; {\bf 933}:L6.

\bibitem{2021ApJ...920...46D}
{Dai} S, {Lu} J, {Wang} C {\em et~al.}
\newblock {On the Circular Polarization of Repeating Fast Radio Bursts}.
\newblock {\em \apj}  2021; {\bf 920}:46.

\bibitem{Zhang2023RMP}
{Zhang} B.
\newblock {The physics of fast radio bursts}.
\newblock {\em Reviews of Modern Physics}  2023; {\bf 95}:035005.

\bibitem{Qiao98}
{Qiao} GJ and {Lin} WP.
\newblock {An inverse Compton scattering (ICS) model of pulsar emission. I.
  Core and conal emission beams}.
\newblock {\em \aap}  1998; {\bf 333}: 172--180.

\bibitem{Xu00}
{Xu} RX, {Liu} JF, {Han} JL {\em et~al.}
\newblock {An Inverse Compton Scattering Model of Pulsar Emission. III.
  Polarization}.
\newblock {\em \apjs}  2000; {\bf 535}: 354--364.

\bibitem{1984ApJS...55..247S}
{Stinebring} DR, {Cordes} JM, {Rankin} JM {\em et~al.}
\newblock {Pulsar polarization fluctuations. I. 1404 MHz statistical
  summaries.}
\newblock {\em \apjs}  1984; {\bf 55}: 247--277.

\bibitem{Singh2024MNRAS.527.2612S}
{Singh} S, {Gupta} Y and {De} K.
\newblock {Single pulse polarization study of pulsars B0950 + 08 and B1642 -
  03: micropulse properties and mixing of orthogonal modes}.
\newblock {\em \mnras}  2024; {\bf 527}: 2612--2623.

\bibitem{Dyks2017MNRAS.472.4598D}
{Dyks} J.
\newblock {The origin of radio pulsar polarization}.
\newblock {\em \mnras}  2017; {\bf 472}: 4598--4617.

\bibitem{Feng2022Sci...375.1266F}
{Feng} Y, {Li} D, {Yang} YP {\em et~al.}
\newblock {Frequency-dependent polarization of repeating fast radio
  bursts{\textemdash}implications for their origin}.
\newblock {\em Science}  2022; {\bf 375}: 1266--1270.

\bibitem{2024A&A...683A.183M}
{Men} Y and {Barr} E.
\newblock {TransientX: A high-performance single-pulse search package}.
\newblock {\em \aap}  2024; {\bf 683}:A183.

\bibitem{PSRCHIVE2004PASA}
{Hotan} AW, {van Straten} W and {Manchester} RN.
\newblock {PSRCHIVE and PSRFITS: An Open Approach to Radio Pulsar Data Storage
  and Analysis}.
\newblock {\em \pasa}  2004; {\bf 21}: 302--309.

\bibitem{PSRCHIVE2010PASA}
{van Straten} W, {Manchester} RN, {Johnston} S {\em et~al.}
\newblock {PSRCHIVE and PSRFITS: Definition of the Stokes Parameters and
  Instrumental Basis Conventions}.
\newblock {\em \pasa}  2010; {\bf 27}: 104--119.

\bibitem{FASTperformance}
{Jiang} P, {Tang} NY, {Hou} LG {\em et~al.}
\newblock {The fundamental performance of FAST with 19-beam receiver at L
  band}.
\newblock {\em Research in Astronomy and Astrophysics}  2020; {\bf 20}:064.

\bibitem{EVN2021ATel14603}
{Marcote} B, {Kirsten} F, {Hessels} JWT {\em et~al.}
\newblock {VLBI localization of FRB 20201124A and absence of persistent
  emission on milliarcsecond scales}.
\newblock {\em The Astronomer's Telegram}  2021; {\bf 14603}: 1.

\bibitem{FASTcommission}
{Jiang} P, {Yue} Y, {Gan} H {\em et~al.}
\newblock {Commissioning progress of the FAST}.
\newblock {\em Science China Physics, Mechanics, and Astronomy}  2019; {\bf
  62}:959502.

\bibitem{Men2019MNRAS}
{Men} YP, {Luo} R, {Chen} MZ {\em et~al.}
\newblock {Piggyback search for fast radio bursts using Nanshan 26 m and
  Kunming 40 m radio telescopes - I. Observing and data analysis systems,
  discovery of a mysterious peryton}.
\newblock {\em \mnras}  2019; {\bf 488}: 3957--3971.

\bibitem{DSPSR2011PASA}
{van Straten} W and {Bailes} M.
\newblock {DSPSR: Digital Signal Processing Software for Pulsar Astronomy}.
\newblock {\em \pasa}  2011; {\bf 28}: 1--14.

\bibitem{PSRCHIVE2012AR&T}
{van Straten} W, {Demorest} P and {Oslowski} S.
\newblock {Pulsar Data Analysis with PSRCHIVE}.
\newblock {\em Astronomical Research and Technology}  2012; {\bf 9}: 237--256.

\bibitem{Dunning2017}
Dunning A, Bowen M, Castillo S {\em et~al.}
\newblock Design and laboratory testing of the five hundred meter aperture
  spherical telescope (fast) 19 beam l-band receiver.
\newblock {\em XXXIInd General Assembly and Scientific Symposium of the
  International Union of Radio Science} (2017) 1--4.

\bibitem{Desvignes2019Sci}
{Desvignes} G, {Kramer} M, {Lee} K {\em et~al.}
\newblock {Radio emission from a pulsar{\textquoteright}s magnetic pole
  revealed by general relativity}.
\newblock {\em Science}  2019; {\bf 365}: 1013--1017.

\bibitem{Schnitzeler2015MNRAS}
{Schnitzeler} DHFM and {Lee} KJ.
\newblock {Rotation measure synthesis revisited.}
\newblock {\em \mnras}  2015; {\bf 447}: L26--L30.

\bibitem{ionfr}
{Sotomayor-Beltran} C, {Sobey} C, {Hessels} JWT {\em et~al.}
\newblock {Calibrating high-precision Faraday rotation measurements for LOFAR
  and the next generation of low-frequency radio telescopes}.
\newblock {\em \aap}  2013; {\bf 552}:A58.

\bibitem{Everett2001ApJ}
{Everett} JE and {Weisberg} JM.
\newblock {Emission Beam Geometry of Selected Pulsars Derived from Average
  Pulse Polarization Data}.
\newblock {\em \apjs}  2001; {\bf 553}: 341--357.

\bibitem{2020Natur.582..351C}
{Chime/Frb Collaboration}, {Amiri} M, {Andersen} BC {\em et~al.}
\newblock {Periodic activity from a fast radio burst source}.
\newblock {\em Nature}  2020; {\bf 582}: 351--355.

\bibitem{NR3}
{Press} W, {Teukolsky} S, {Vetterling} WT {\em et~al.}
\newblock {\em Numerical Recipes 3rd Edition: The Art of Scientific Computing}
  (Cambridge University Press, New York, NY, USA), 3 edition.

\bibitem{Dulk1985ARA&A}
{Dulk} GA.
\newblock {Radio emission from the sun and stars.}
\newblock {\em \araa}  1985; {\bf 23}: 169--224.

\bibitem{2016A&A...587A...3G}
{Girard} JN, {Zarka} P, {Tasse} C {\em et~al.}
\newblock {Imaging Jupiter's radiation belts down to 127 MHz with LOFAR}.
\newblock {\em \aap}  2016; {\bf 587}:A3.

\bibitem{2014JGRA..119.9508C}
{Clarke} TE, {Higgins} CA, {Skarda} J {\em et~al.}
\newblock {Probing Jovian decametric emission with the long wavelength array
  station 1}.
\newblock {\em Journal of Geophysical Research (Space Physics)}  2014; {\bf
  119}: 9508--9526.

\bibitem{2017A&A...604A..17M}
{Marques} MS, {Zarka} P, {Echer} E {\em et~al.}
\newblock {Statistical analysis of 26 yr of observations of decametric radio
  emissions from Jupiter}.
\newblock {\em \aap}  2017; {\bf 604}:A17.

\bibitem{Han1998MNRAS}
{Han} JL, {Manchester} RN, {Xu} RX {\em et~al.}
\newblock {Circular polarization in pulsar integrated profiles}.
\newblock {\em \mnras}  1998; {\bf 300}: 373--387.

\bibitem{2018MNRAS.474.4629J}
{Johnston} S and {Kerr} M.
\newblock {Polarimetry of 600 pulsars from observations at 1.4 GHz with the
  Parkes radio telescope}.
\newblock {\em \mnras}  2018; {\bf 474}: 4629--4636.

\bibitem{2022PASA...39...27S}
{Spiewak} R, {Bailes} M, {Miles} MT {\em et~al.}
\newblock {The MeerTime Pulsar Timing Array: A census of emission properties
  and timing potential}.
\newblock {\em \pasa}  2022; {\bf 39}:e027.

\bibitem{psrcat}
{Manchester} RN, {Hobbs} GB, {Teoh} A {\em et~al.}
\newblock {The Australia Telescope National Facility Pulsar Catalogue}.
\newblock {\em \aj}  2005; {\bf 129}: 1993--2006.

\bibitem{handbook}
{Lorimer} DR and {Kramer} M.
\newblock {\em {Handbook of Pulsar Astronomy}} (Cambridge University Press, New
  York, NY, USA).

\bibitem{Michilli2018Natur}
{Michilli} D, {Seymour} A, {Hessels} JWT {\em et~al.}
\newblock {An extreme magneto-ionic environment associated with the fast radio
  burst source FRB 121102}.
\newblock {\em \nat}  2018; {\bf 553}: 182--185.

\bibitem{Gajjar2018ApJ}
{Gajjar} V, {Siemion} APV, {Price} DC {\em et~al.}
\newblock {Highest Frequency Detection of FRB 121102 at 4-8 GHz Using the
  Breakthrough Listen Digital Backend at the Green Bank Telescope}.
\newblock {\em \apjs}  2018; {\bf 863}:2.

\bibitem{2021RNAAS...5...17F}
{Faber} JT, {Gajjar} V, {Siemion} APV {\em et~al.}
\newblock {Re-analysis of Breakthrough Listen Observations of FRB 121102:
  Polarization Properties of Eight New Spectrally Narrow Bursts}.
\newblock {\em Research Notes of the American Astronomical Society}  2021; {\bf
  5}:17.

\bibitem{Hilmarsson2021ApJ}
{Hilmarsson} GH, {Michilli} D, {Spitler} LG {\em et~al.}
\newblock {Rotation Measure Evolution of the Repeating Fast Radio Burst Source
  FRB 121102}.
\newblock {\em \apjl}  2021; {\bf 908}:L10.

\bibitem{2023MNRAS.526.3652K}
{Kumar} P, {Luo} R, {Price} DC {\em et~al.}
\newblock {Spectropolarimetric variability in the repeating fast radio burst
  source FRB 20180301A}.
\newblock {\em \mnras}  2023; {\bf 526}: 3652--3672.

\bibitem{2023MNRAS.526.2039H}
{Hewitt} DM, {Hessels} JWT, {Ould-Boukattine} OS {\em et~al.}
\newblock {Dense forests of microshots in bursts from FRB 20220912A}.
\newblock {\em \mnras}  2023; {\bf 526}: 2039--2057.

\end{thebibliography}

\end{document}